\documentclass[aps,prb,reprint,superscriptaddress]{revtex4-2}

\usepackage{amsmath}
\usepackage{amssymb}
\usepackage{braket}
\usepackage{bm}
\usepackage{graphicx}
\usepackage{hyperref}
\usepackage{hyphenat}
\usepackage{siunitx}

\graphicspath{{../figures/}}
\hypersetup{allcolors=black}
\hypersetup{colorlinks=true}
\hypersetup{urlcolor=blue}

\newcommand{\dd}{\text{d}}
\newcommand{\s}{\sigma}
\newcommand{\kk}{{\bm k}}
\newcommand{\rr}{{\bm \rho}}
\newcommand{\rp}{\hat{\bm \rho}}
\newcommand{\pa}{\parallel}

\setcitestyle{numbers,square,citesep={,\kern-.24em}}

\begin{document}

\title{Nonlinear transverse current on $\mathcal{C}_{3v}$-symmetric honeycomb
lattice}

\author{Francesco Peronaci}
\affiliation{Max Planck Institute for the Physics of Complex Systems, Dresden
01187, Germany}
\email{francesco.peronaci@gmail.com}
\author{Takashi Oka}
\affiliation{Max Planck Institute for the Physics of Complex Systems, Dresden
01187, Germany}
\affiliation{Institute for Solid State Physics, University of Tokyo, Kashiwa
277-8581, Japan}

\date{\today}

\begin{abstract}
At nonlinear orders in the electric field, the vanishing of the Hall
conductivity does not prevent the nonlinear component of the current from
being transverse for selected field directions. We study electrons on
$\mathcal{C}_{3v}$-symmetric honeycomb lattice for which the Hall conductivity
vanishes at first and second order. Nevertheless, the second-order current
component is transverse for fields perpendicular to the three mirror lines.
The $\mathcal{C}_{3v}$ symmetry constrains the first-order and second-order
conductivity tensors to have only one independent component each, which we
calculate using the quantum kinetic equation. In linearly\hyp{}polarized
oscillating field, the current has a zero\hyp{}frequency component switching
sign upon $\pi/2$-rotation of the polarization angle.
\end{abstract}

\maketitle

The description of the quantum Hall effect in terms of the electronic Berry
phase is a milestone of modern solid state physics\,\cite{ Xiao2010}. At linear
order in the applied field, the Hall conductivity is related to the Berry
curvature and vanishes by time-reversal symmetry\,\cite{ Nagaosa2010}. At
second order, it is induced by the Berry-curvature dipole and can be finite
also in time\hyp{}reversal\hyp{}invariant systems, provided a low crystal
symmetry\,\cite{ Sodemann2015}. Specifically, it requires broken inversion
symmetry and at most one mirror symmetry and has been indeed detected in
layered transition\hyp{}metal dichalcogenides with one mirror plane\,\cite{
Ma2019, Kang2019}.

In this work we consider electrons on the honeycomb lattice with a
site-potential imbalance, whose symmetry point group $\mathcal{C}_{3v}$
features a $2\pi/3$-rotation axis and three mirror lines. Since the system has
also time\hyp{}reversal symmetry, the Hall conductivity vanishes at first and
second order. Nevertheless, the second-order current component is reminiscent
of a Hall current, since it is transverse for fields perpendicular to the
mirror lines. Within the quantum\hyp{}kinetic framework, we numerically
calculate the conductivity tensor up to second order and discuss its
dependence on site\hyp{{}potential imbalance and relaxation time. Finally, we
calculate the response to linearly\hyp{}polarized oscillating field and its
zero\hyp{}frequency component, which originates from the second\hyp{}order
response and switches sign upon $\pi/2$\hyp{}rotation of the polarization
angle.

\section{\label{sec1}Nonlinear transverse current in systems with vanishing
Hall conductivity}

Let us start by illustrating in simple terms the unique property of nonlinear
current responses, namely they can be transverse, meaning perpendicular to the
applied field, even with vanishing Hall conductivity. Let us consider a
two\hyp{}dimensional system and assume the electric\hyp{}current density
$j_\alpha$ ($\alpha=x,y$) admits a power-series expansion in the electric field
$E_\alpha$, which up to second order reads
\begin{equation}
\label{eq_j}
j_\alpha = \sigma_{\alpha\beta} E_\beta + \chi_{\alpha\beta\gamma} E_\beta
E_\gamma.
\end{equation}
The Hall conductivity at each order is the antisymmetric part of the
conductivity, $\epsilon_{\alpha\beta} \epsilon_{\zeta\eta} \sigma_{\zeta\eta}$
and $\epsilon_{\alpha\beta} \epsilon_{\zeta\eta} \chi_{\zeta\eta\gamma}$, hence
the symmetry constraints on the Hall effect\,\cite{ Nandy2019, Matsyshyn2019,
Note3}. The Hall current is therefore by definition transverse and
dissipationless, namely it does not contribute to the power
\begin{equation}
\label{eq_p}
j_\alpha E_\alpha = \sigma_{\alpha\beta} E_\alpha E_\beta +
\chi_{\alpha\beta\gamma} E_\alpha E_\beta E_\gamma.
\end{equation}
The two terms on the right-hand-side of Eq.\,\eqref{eq_p} have a remarkable
difference: the first is a quadratic form and vanishes if and only if
$\sigma_{\alpha\beta}$ is antisymmetric; the second is a cubic form and the
antisymmetry of $\chi_{\alpha\beta\gamma}$ in $\alpha$ and $\beta$ is
\emph{sufficient} but not \emph{necessary} for it to vanish. Indeed, there is
in general at least one field direction which makes the second term on the
right-hand-side of Eq.\,\eqref{eq_p} to vanish, meaning that the second-order
component in Eq.\,\eqref{eq_j} is transverse\,\cite{Note2}. In other words, at
(and only at) nonlinear orders in the applied electric field, the nonlinear
component of the current is transverse for selected field directions, no matter
the vanishing of the Hall effect.

\footnotetext[2]{To see this, plug $E_y/E_x=m$ into Eq.\,\eqref{eq_p}. The
first term is proportional to $\sigma_{yy}m^2 + (\sigma_{xy}+\sigma_{yx})m +
\sigma_{xx}$ which has no real zero ($\sigma_{\alpha\beta}$ positive definite);
the second term is proportional to $\chi_{yyy}m^3 + (\chi_{yyx} + \chi_{yxy} +
\chi_{xyy})m^2 + (\chi_{xxy} + \chi_{xyx} + \chi_{yxx})m + \chi_{xxx}$ with
at least one real zero.}

\footnotetext[3]{Note that $\chi_{\alpha\beta\gamma}$ is symmetric in $\beta$
and $\gamma$ by construction.}

\section{\label{sec2}Second-order current in $\mathcal{C}_{3v}$-symmetric
systems}

We now turn to the case of $\mathcal{C}_{3v}$ symmetry, which applies to the honeycomb
lattice with site\hyp{}potential imbalance, see Fig.\,\ref{fig_1}. To derive
the symmetry constraints, we fix a choice of coordinate axes -- $\hat x$
parallel to one of the mirror lines -- and impose the symmetries $\sigma_v'$
and $C_3$, namely the reflection with respect to the $x$ axis and the
$2\pi/3$-rotation. The resulting constraints read\,\cite{Note1}
\begin{subequations}
\label{eq_4}
\begin{gather*}
\begin{align}
\sigma_{xy} = \sigma_{yx} &= 0, \\
\sigma_{xx} = \sigma_{yy} &\equiv \sigma,
\end{align} \\
\begin{align}
\chi_{xxy} = \chi_{xyx} = \chi_{yxx} = \chi_{yyy} &= 0, \\
\label{eq_3b}
-\chi_{xxx} = \chi_{xyy} = \chi_{yxy} = \chi_{yyx} &\equiv - \chi.
\end{align}
\end{gather*}
\end{subequations}
Thus the entire response up to second order depends only on two coefficients
$\sigma$ and $\chi$ (not to be confused with the tensors
$\sigma_{\alpha\beta}$, $\chi_{\alpha\beta\gamma}$). In particular,
Eq.\,\eqref{eq_3b} states that the second-order non-diagonal components with
odd number of ``$x$-indices'' are equal to each other but need not be
vanishing. The latter fact relates to the broken $x$\hyp{}axis\hyp{}reversal
symmetry (reflection with respect to the $y$ axis) by the site-potential
imbalance, see Fig.\,\ref{fig_1}. The meaning of $\chi_{xyy}$ is particularly
clear as it gives a response along $\hat x$ to a field along $\hat y$. Note
that the Hall conductivity vanishes at first ($\epsilon_{\alpha\beta}
\sigma_{\alpha\beta} =0$) and second order ($\epsilon_{\alpha\beta}
\chi_{\alpha\beta\gamma} =0$) consistently with well known symmetry
constraints\,\cite{ Nandy2019}.

\begin{figure}
\includegraphics[width=0.8\columnwidth]{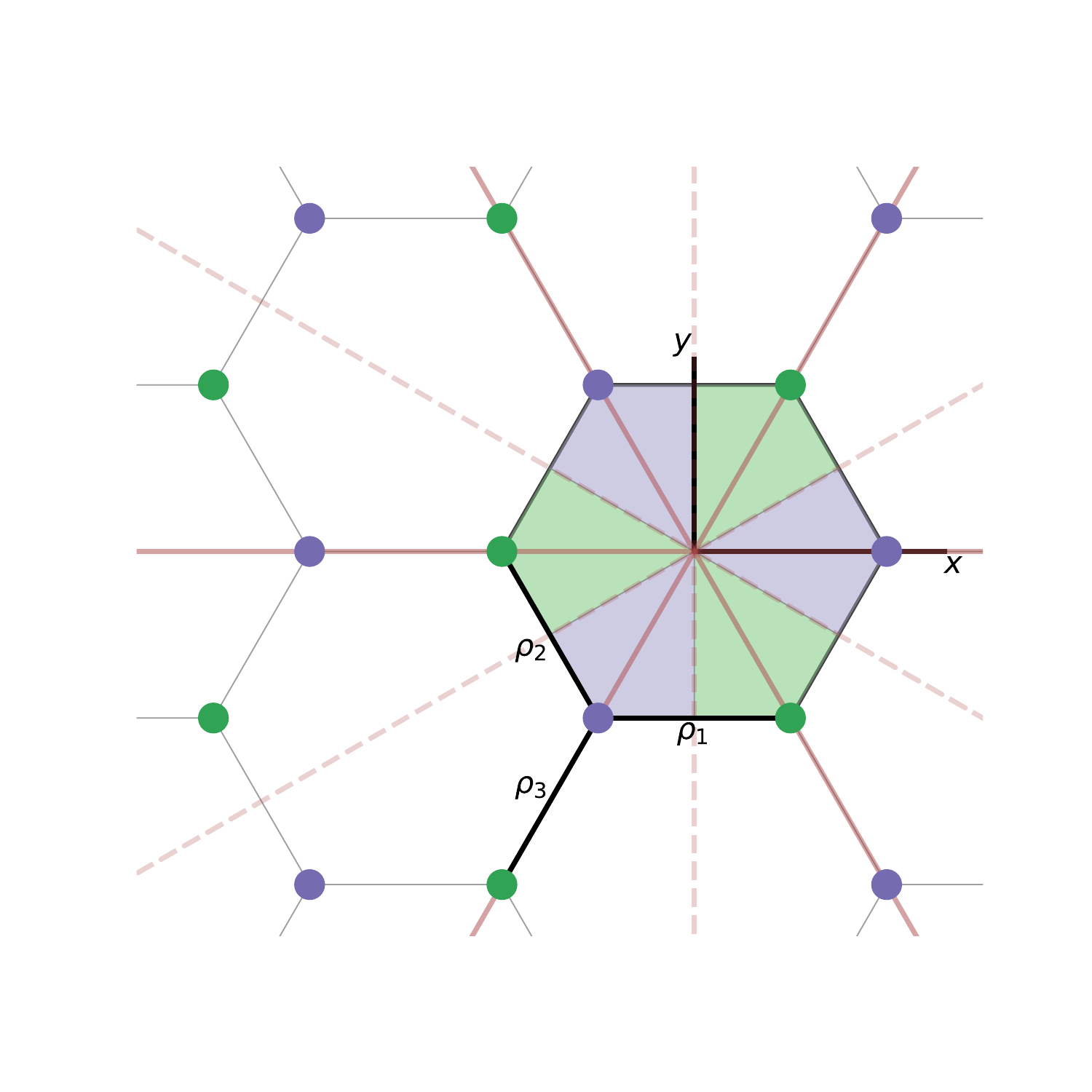}
\caption{\label{fig_1}$\mathcal{C}_{3v}$-symmetric honeycomb lattice. Sites on
sublattice $A$ (blue) and $B$ (green) have potential imbalance $\pm m$. The
symmetry point group (with respect to, e.g., the center of each hexagon)
include rotations of $2\pi/3$ and $4\pi/3$ and three reflections (with respect
to the solid orange lines). For $m=0$ appear three additional mirror lines
(dashed orange lines).}
\end{figure}

Plugging Eqs.\,\eqref{eq_4} into Eq.\,\eqref{eq_j} yields the current up to
second order in the field $\bm E = (E_x,E_y) = E(\cos\theta,\sin\theta)$ of a
system with $\mathcal{C}_{3v}$ symmetry:
\begin{subequations}
\label{eq_2}
\begin{align}
j_x &= \sigma E_x + \chi (E_x^2-E_y^2), \\
j_y &= \sigma E_y -2\chi E_x E_y.
\end{align}
\end{subequations}
The second-order component forms an angle $-2\theta$ with $\hat x$, hence it is
transverse for $\theta=\pi/6+n\pi/3$ and longitudinal for $\theta=n\pi/3$ ($n$
integer). In other words, it is transverse or longitudinal for fields
perpendicular or parallel to the three mirror lines, which on honeycomb
lattice coincide with the three bond directions $\hat
\rr_n=(\cos\phi_n,\sin\phi_n)$ with $\phi_n=2\pi n/3$ ($n=1,2,3)$, see
Fig.\,\ref{fig_1}.

Equations\,\eqref{eq_2} can be written in the invariant fashion which does not
depend on the choice of coordinate axes and makes the $\mathcal{C}_{3v}$
symmetry manifest together with the direction of the second-order current:
\begin{equation}
\label{eq_5}
\bm{j} = \sigma \bm E + (4/3)\chi \sum_{(i,j,k)\in P_3}
(\bm E\cdot\hat\rr_i) (\bm E\cdot \hat\rr_j) \hat\rr_k,
\end{equation}
where $P_3 = \{(1,2,3),(2,3,1),(3,1,2)\}$. Furthermore, we can express the
conductivity tensors as $\sigma_{\alpha\beta} = \sigma\delta_{\alpha\beta}$ and
$\chi_{\alpha\beta\gamma} = (4/3)\chi\ \sum_{(i,j,k)\in P_3} \hat\rho_{i\alpha}
\hat\rho_{j\beta} \hat\rho_{k\gamma}$ which make evident the symmetry of these
tensors, hence the vanishing of the Hall effect. The dissipated power reads
\begin{equation}
\bm{j}\cdot\bm{E} = \sigma |\bm E|^2 + 4\chi (\bm E\cdot\hat\rr_1) (\bm
E\cdot\hat\rr_2)(\bm E\cdot\hat\rr_3),
\end{equation}
and it is clear that the second-order current component is dissipationless for
fields perpendicular to the three bonds.

\section{\label{sec3}Model and methods}

To proceed with the calculation of conductivity tensors and nonlinear
transverse current, we consider the tight-binding Hamiltonian on honeycomb
lattice\,\cite{ Neto2009}
\begin{gather}
\label{eq_h} H = \sum\nolimits_\kk\Psi_\kk^\dagger H_\kk\Psi_\kk, \\
H_\kk = (\text{Re}f_\kk)t_h\sigma_1 + (\text{Im}f_\kk)t_h\sigma_2 + m\sigma_3,
\\ f_k = \exp(i \kk\cdot\rr_1) + \exp(i \kk\cdot\rr_2) + \exp(i \kk\cdot\rr_3).
\end{gather}
Here $\Psi_\kk=(c_{\kk A},c_{\kk B})^T$ where $c_{\kk A(B)}$ is the lattice
Fourier transform of the electronic annihilation operator on the sublattice $A(B)$,
$\sigma_i$ ($i=1,2,3$) are Pauli matrices, $t_h$ is the nearest-neighbor
hopping, $m$ is the site-potential imbalance and $\bm \rho_n=a\rp_n$
with $a$ nearest-neighbor distance.

The density matrix $\rho_\kk^{ab} = \braket{ \Psi_\kk^{b\dagger} \Psi_\kk^a }$
($a,b=1,2$) evolves in time according to the quantum kinetic equation
\begin{equation}
\label{eq_qke}
\dot\rho_\kk=i\hbar^{-1}[\rho_\kk,H_{\kk+e\bm A(t)/\hbar}]
-\tau^{-1}(\rho_\kk-\rho_{\kk+e\bm A(t)/\hbar}^0).
\end{equation}
The first term on the right-hand side of Eq.\,\eqref{eq_qke} gives the unitary
evolution with Hamiltonian with shifted crystal momentum, where $\bm A(t)=-\bm
Et$ is the gauge field and $\bm E$ the electric field. The second term provides
dissipation with relaxation time $\tau$, $\rho_\kk^0 =
[1+\exp(H_\kk/k_BT)]^{-1}$ is the equilibrium density matrix at a temperature
$T=0$. The current density is $\bm{j} = (2\pi^2)^{-1}\int \text{d}^2
\kk\text{Tr}(\rho_\kk e\bm v_{\kk+e\bm A(t)/\hbar})$ where $\bm v_\kk =
\hbar^{-1} \nabla_\kk H_\kk$, the integral is over the Brillouin zone and both
spin directions are taken into account.

The system is a band insulator with gap $\Delta_g = 2m$ for $m\ne0$ and a
semimetal for $m=0$. Equation\,\eqref{eq_j} is typically used only for metallic
systems, see e.\,g.\ Ref.\,\cite{ Nandy2019}. To use it also for $m\ne0$ is
acceptable for a certain range of $\tau$ and $E$, as we discuss again below and
as ultimately confirmed by the numerical integration of Eq.\,\eqref{eq_qke} in
Sec.\,\ref{sec4}.

The quantum kinetic equation with relaxation\hyp{}time approximation,
Eq.\,\eqref{eq_qke}, has been recently used in similar setups, for instance to
study the light-induced anomalous Hall effect in graphene\,\cite{Sato2019}, see
also Refs.\,\cite{Oka2009, Mciver2020}, and the electron-hole pair creation in
rotating electric fields\,\cite{Takayoshi2021}.

\section{\label{sec4}First-order and second-order conductivity coefficients}

We present here the numerical calculation of the time evolution in static
electric field and, from the steady-state current, we extract the first-order
$\sigma$ and second-order $\chi$ conductivity coefficients, discussing the
dependence on site-potential imbalance $m$ and relaxation time $\tau$.

\begin{figure}
\includegraphics[width=0.49\columnwidth]{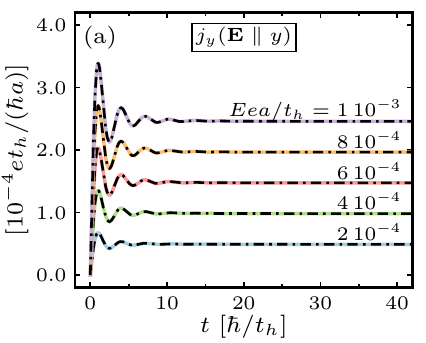}
\includegraphics[width=0.49\columnwidth]{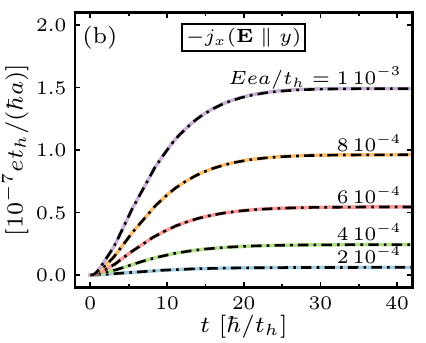}
\includegraphics[width=0.49\columnwidth]{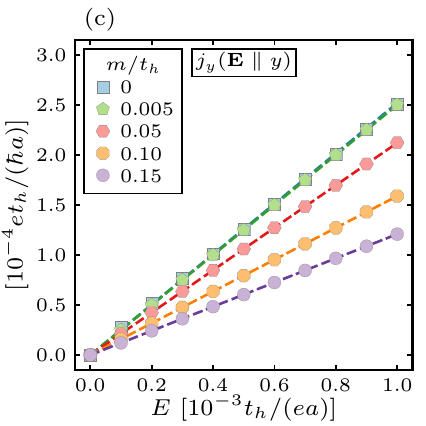}
\includegraphics[width=0.49\columnwidth]{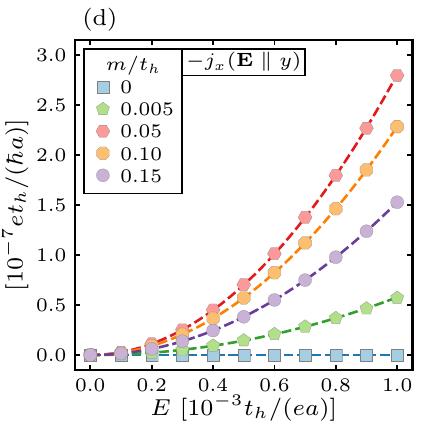}
\caption{\label{fig_2}Top panels: Time evolution of (solid) (a) longitudinal
current $j_y(\bm E \pa \hat y)$ and (b) transverse current $j_x(\bm E \pa \hat
y)$; with (dash-dotted) symmetrized longitudinal current $j_x(\bm E \pa \hat
x)$ according to Eqs.\,\eqref{eq_6}; $m/t_h=0.015$, $\tau t_h/\hbar=5$. Bottom
panels: Steady-state (markers) (c) longitudinal current $j_y(\bm E \pa \hat y)$
and (d) transverse current $j_x(\bm E \pa \hat y)$; with (dashed) respectively
linear and quadratic fits; $\tau t_h/\hbar=5$.}
\end{figure}

In Fig.\,\ref{fig_2}(a),(b) we plot the time evolution of $j_y(\bm E \pa \hat
y)$ and $j_x(\bm E \pa \hat y)$, longitudinal and transverse components of the
current. In agreement with our expectations based on Eqs.\,\eqref{eq_2}, for
$m\ne0$ and field along $\hat y$ (perpendicular to a mirror line) the system
sustains a transverse current invariant upon field reversal ($\bm E \rightarrow - \bm E$), as we have also checked (not shown). To further substantiate
Eqs.\,\eqref{eq_2}, we derive opportune combinations of longitudinal currents
$j_x(\bm E \pa \pm \hat x)$ in fields of the same amplitude along $\pm\hat x$,
which relate to the current in field along $\hat y$ as
\begin{subequations}
\label{eq_6}
\begin{align}
\label{eq_6a}
\frac{j_x(E,0)-j_x(-E,0)}{2} &= j_y(0,E), \\
\label{eq_6b}
\frac{j_x(E,0)+j_x(-E,0)}{2} &= -j_x(0,E).
\end{align}
\end{subequations}
The left\hyp{}hand\hyp{}side of Eqs.\,\eqref{eq_6}, derived from
Eqs.\,\eqref{eq_2}, is also plotted in Fig.\,\ref{fig_2}(a),(b) (dash-dotted),
supporting the validity of the power-series expansion Eq.\,\eqref{eq_j} and of
the consequent Eqs.\,\eqref{eq_2} in this range of parameters.

The steady-state current versus the field amplitude is plotted in
Fig.\,\ref{fig_2}(c),(d). Longitudinal and transverse currents are linear and
quadratic, respectively. At fixed $E$, the former decreases with~$m$, while the
latter vanishes for $m=0$, then increases with $m$ and finally also decreases.

\begin{figure}
\includegraphics[width=0.49\columnwidth]{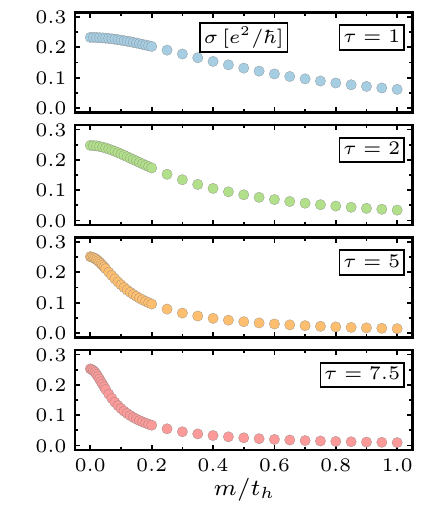}
\includegraphics[width=0.49\columnwidth]{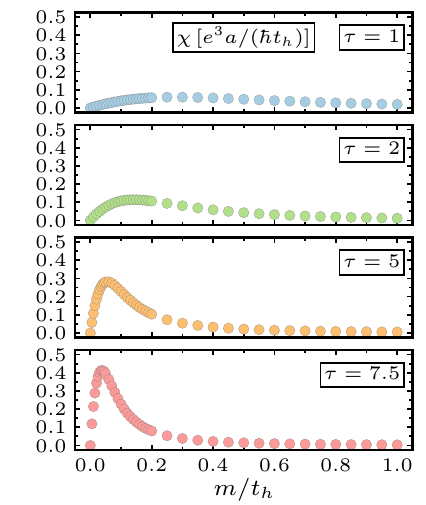}
\includegraphics[width=0.49\columnwidth]{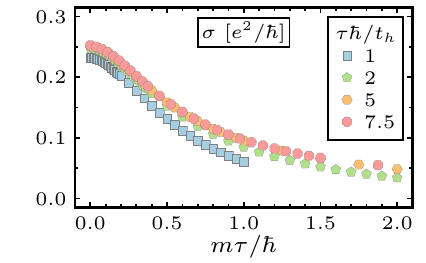}
\includegraphics[width=0.49\columnwidth]{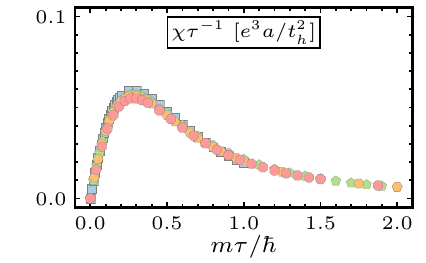}
\caption{\label{fig_3}Top panels: First\hyp{}order (left) and second\hyp{}order
(right) conductivity coefficients $\sigma$, $\chi$ versus site-potential
imbalance $m/t_h$ and for varying relaxation time $\tau$. $\sigma$ is
monotonous decreasing to zero at large $m$, $\chi$ vanishes at $m=0$, is
maximum at $m^*$, and finally goes to zero at large $m$. Bottom panels: curve
collapse with $\sigma\rightarrow\sigma$, $\chi\rightarrow \chi\tau^{-1}$,
$m\rightarrow m\tau$ showing the scaling relations $\sigma(m=0)\sim\tau^0$,
$\chi(m=m^*)\sim\tau^1$, $m^*\sim\tau^{-1}$.}
\end{figure}

To extract the coefficients $\sigma$ and $\chi$, we fit longitudinal and
transverse steady-state currents as $j_y(\bm E \pa \hat y) = \sigma E$ and
$j_x(\bm E \pa \hat y) = -\chi E^2$ cf.\ Eqs.\,\eqref{eq_2}. The fitting lines
are plotted in Fig.\,\ref{fig_2}(c),(d) (dashed) and the coefficients in
Fig.\,\ref{fig_3} (top panels) versus $m$ and for varying $\tau$.

The numerical result shows that, for each $m$, $\tau$, Eqs.\,\eqref{eq_2} are
valid at small enough $E$\,\cite{Note1}. Decreasing $m$ or $\tau$ requires
smaller $E$ in order to extract $\sigma$ and $\chi$. Note that, indeed, for
small $E$ and $\tau$ the density matrix expands in a double
power\hyp{}series\,\cite{ Nandy2019} leading in turn to a power\hyp{}series for
the current. Specifically, here we use $Eea/t_h<0.01$ for $m/t_h\ge 0.25$
and/or $\tau t_h/\hbar=1,2$; $Eea/t_h<0.001$ for $m/t_h<0.25$ and $\tau
t_h/\hbar=5,7.5$.

\footnotetext[1]{See Supplemental Materials.}

For $m=0$, $\sigma$ is maximum and $\chi$ vanishes. Indeed in this limiting
case the band gap closes and three mirror lines appear, which further constrain
the components of the conductivity $\chi_{\alpha\beta\gamma}$. Increasing $m$,
the band gap opens and widens. Accordingly, $\sigma$ decreases while $\chi$ is
instead non\hyp{}monotonous: first it becomes finite as a consequence of the
broken $x$-axis reversal symmetry, then reaches a maximum for $m=m^*$and
finally also decreases. In the limit of large $m$ the band structure consists
of two isolated bands -- one full the other empty -- and at all orders the
conductivity trivially vanishes.

Increasing $\tau$ the conductivity coefficients decrease for all $m$ except for
their maxima $\sigma(m=0)$ and $\chi(m=m^*)$ which are, respectively, constant
and increasing with $\tau$. Moreover, $m^*$ which makes $\chi$ maximum
decreases with $\tau$. Guided by these observations, in Fig.\,\ref{fig_3}
(bottom panels) we plot rescaled conductivity coefficients $\sigma\tau^{0}$ and
$\chi\tau^{-1}$ versus rescaled site-potential $m\tau^1$. The curve collapse
shows the validity of the approximate scaling relations
$\sigma(m=0)\sim\tau^0$, $\chi(m=m^*)\sim\tau^1$ and $m^*\sim\tau^{-1}$.

For $m=0$ we get $\sigma\approx0.25\,e^2/\hbar$ not far from the theoretical
value for two\hyp{}dimensional massless Dirac electrons $4e^2/(\pi h) \approx
0.2\,e^2/\hbar$\,\cite{ Neto2009}. Moreover $\chi\sim \tau^1$ has the
same scaling as the nonlinear Hall conductivity in Refs.\,\cite{ Sodemann2015,
Nandy2019}.

\section{\label{sec5}Zero-frequency current component in oscillating field}

\begin{figure}
\includegraphics[width=0.49\columnwidth]{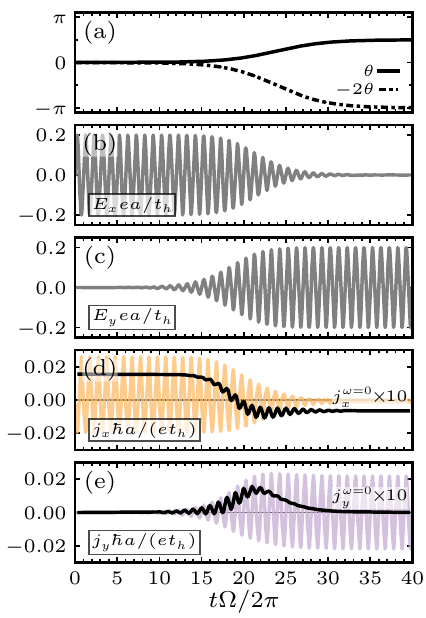}
\includegraphics[width=0.49\columnwidth]{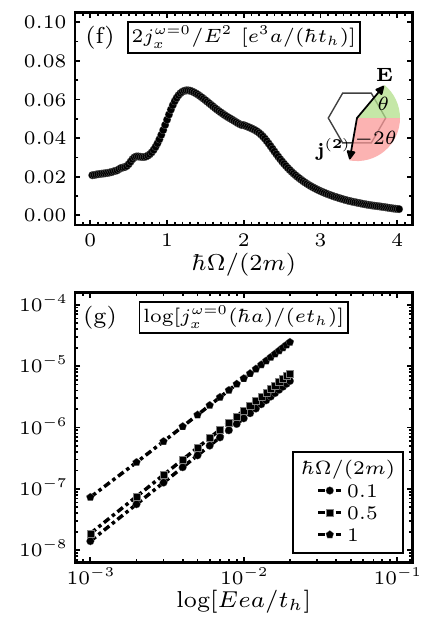}
\caption{\label{fig_6}(a) Modulation of polarization angle of applied field
$\bm E=E\cos(\Omega t) (\cos\theta,\sin\theta)$ and of second\hyp{}order
response $-2\theta$ [cf.\ inset panel (f)]. (b),(c)~Corresponding field
components. (d),(e)~Time evolution of current (color) and of
zero\hyp{}frequency component (black); $m/t_h=0.5$, $\tau t_h/\hbar=5$,
$\hbar\Omega/t_h=0.1$. (f)~Zero-frequency component $2j_x^{\omega=0}/E^2$ with
$\bm E ea/t_h= 0.1\,\hat x$ versus field frequency $\hbar\Omega/(2m)$.
(g)~$j_x^{\omega=0}(\bm E \pa \hat x)$ versus field amplitude for varying field
frequency.}
\end{figure}

Let us now consider an oscillating applied field. Since the second-order
current response is invariant under field reversal [$\bm E \rightarrow - \bm
E$, cf.\ Eqs.\,\eqref{eq_2}], we expect it to have a finite time average,
namely a zero-frequency component. Before presenting the numerical result, we
gain insight into the response to a linearly-polarized oscillating field $\bm E
= E\cos(\Omega t) (\cos\theta,\sin\theta)$ by substitution in
Eqs.\,\eqref{eq_2}:
\begin{subequations}
\label{eq_je}
\begin{align}
j_x &=\sigma E\cos(\Omega t)\cos\theta+\chi E^2[\cos(\Omega t)]^2\cos2\theta, \\
j_y &=\sigma E\cos(\Omega t)\sin\theta-\chi E^2[\cos(\Omega t)]^2\sin2\theta,
\end{align}
\end{subequations}
Since Eqs.\,\eqref{eq_2} are formulated for static electric field, we expect
Eqs.\,\eqref{eq_je} to be approximately valid only for small frequency
$\Omega$. The time average, that is the zero\hyp{}frequency ($\omega=0$)
current component, reads
\begin{subequations}
\label{eq_jee}
\begin{align}
j_x^{\omega=0} &= (\chi E^2/2)\cos2\theta, \\
j_y^{\omega=0} &=-(\chi E^2/2)\sin2\theta.
\end{align}
\end{subequations}
Thus indeed a zero\hyp{}frequency component originates from the second-order
response and forms the same angle $-2\theta$ with $\hat x$. Besides being
transverse for $\theta=\pi/6+n\pi/3$ and longitudinal for $\theta=n\pi/3$ ($n$
integer), as discussed above, this also implies that it reverses upon
$\pi/2$-rotation of the applied field, since $\theta\rightarrow \theta + \pi/2$
implies $2\theta\rightarrow 2\theta +\pi$.

To substantiate these insights, we consider a field with polarization angle
modulated in time between $0$ and $\pi/2$ as $\theta(t) = \frac{\pi}{4}
[1+\tanh[0.2(t \Omega/2\pi-20)]]$, see Fig.\,\ref{fig_6}(a)-(c). The numerical
result is plotted in Fig.\,\ref{fig_6}(d),(e). The time evolution (color line)
is dominated by the linear terms in Eqs.\,\eqref{eq_je} which yield a current
component in the same direction of the applied field and with same frequency.
On top of this, the nonlinear terms are only appreciable in $j_x$ when
$\theta=\pi/2$, see Fig.\,\ref{fig_6}(d). Also plotted in
Fig.\,\ref{fig_6}(d),(e) is the zero\hyp{}frequency component (black
line) calculated as the running time average over the period of the field
$\bm j^{\omega=0}(t) = \Omega/2\pi \int\dd{t'}\,\bm j(t')$. The result is
in qualitative agreement with Eqs.\,\eqref{eq_jee} and in particular $\bm
j^{\omega=0}$ switches sign upon a $\pi/2$-rotation of the polarization angle.

In Fig.\,\ref{fig_6}(f) we plot $2j_x^{\omega=0}/E^2$ with $\bm E \pa \hat x$
($\theta=0$). From Eqs.\,\eqref{eq_jee} we expect this combination to
approach $\chi$ at small field frequency, which is indeed the case since in
Sec.\,\ref{sec4} for $m/t_h=0.5$, $\tau t_h/\hbar=5$ we have $\chi \approx
0.028$. Increasing the field frequency, $j_x^{\omega=0}$ increases and reaches
a maximum for $\hbar\Omega \approx 2m = \Delta_g$ decreasing then to zero. In
Fig.\,\ref{fig_6}(g) we plot $j_x^{\omega=0}$ with $\theta=0$ in log-log
scale versus field amplitude and for varying field frequency.

\section*{Conclusions}

At each order in the field, the Hall effect contributes a transverse current
\emph{no matter} the field direction. At nonlinear orders, however, also the
remaining current (non-Hall) can be transverse for \emph{selected} field
directions. On $\mathcal{C}_{3v}$\hyp{}symmetric honeycomb lattice, the
second\hyp{}order current is transverse for fields perpendicular to each mirror
line, in spite of the vanishing Hall conductivity.

Integration of the quantum kinetic equation yields the second\hyp{}order
conductivity, which scales as $\chi\sim\tau^{1}$ and is maximum for
site\hyp{}potential imbalance $m=m^*\sim\tau^{-1}$.

In linearly\hyp{}polarized oscillating field the current has a
zero\hyp{}frequency component originating from the second-order response, which
switches sign upon $\pi/2$-rotation of the polarization angle.

\begin{acknowledgments}
FP is grateful to I.\ Sodemann for valuable discussions and to S.\ Kitamura,
S.\ Takayoshi, C.\ Danieli for fruitful interactions. TO was supported by JST
CREST Grant No.\ JPMJCR19T3 and JST ERATO-FS Grant No.\ JPMJER2105, Japan.
\end{acknowledgments}

\bibliography{main}

\clearpage
\section*{Supplemental Materials}

\subsection*{\label{appA}Point group $\mathcal{C}_{3v}$ and symmetry
constraints on conductivity tensors}

The honeycomb lattice with different site potential $\pm m$ on the two
sublattices has symmetry point group $\mathcal{C}_{3v}$ with respect to either
the center of each hexagon or each of its vertices, see Fig.\,\ref{fig_1}. The
group has order $g=6$ (number of elements) and is composed of the symmetry
operations $\{E,C_3,C_3^2,\s_v',\s_v'',\s_v'''\}$ where $E$ is the identity,
$C_3$ is the $2\pi/3$-rotation and $\sigma_v$'s are reflections with respect to
the lines parallel to the sides of the hexagon, see Table\,\ref{tab1} for the
multiplication rules. This is a subgroup of $\mathcal{C}_{6v}$, obtained for
$m=0$, which has the additional symmetries
$\{C_2,C_6,C_6^5,\s_d',\s_d'',\s_d'''\}$ with $\sigma_d$'s reflections with
respect to the lines perpendicular to the sides of the hexagon.

The symmetry operations of the group are divided in three classes $\{E\}$,
$\{C_3,C_3^2\}$, $\{\s_v',\s_v'',\s_v'''\}$ ($A$, $B$ are in the same class if
$B=CAC^{-1}$ for some $C$). Since constraints imposed by symmetry operations
of the same class are not independent, it suffices to take only one operation
per class, for instance $\sigma_v'$ and $C_3$ with representations
\begin{equation}
\label{app_o}
\sigma_v' = \begin{pmatrix} 1 & 0 \\ 0 & -1 \end{pmatrix}, \qquad
C_3 = \frac{1}{2} \begin{pmatrix} -1 & -\sqrt{3} \\ \sqrt{3} & -1 \end{pmatrix}.
\end{equation}

In two dimensions, a rotation or reflection symmetry operation $O$ with
representation $O_{\alpha\beta}$ acts on the conductivity tensors
$\sigma_{\alpha\beta}$ and $\chi_{\alpha\beta\gamma}$ as ($O^{-1}=O^T$)
\begin{subequations}
\begin{gather}
\sigma_{\alpha\beta} \rightarrow \tilde \sigma_{\alpha\beta} = O_{\alpha m}
O_{\beta n} \sigma_{mn}, \\
\chi_{\alpha\beta\gamma} \rightarrow \tilde \chi_{\alpha\beta\gamma} = O_{\alpha m} O_{\beta n} O_{\gamma p} \chi_{mnp}.
\end{gather}
\end{subequations}
If $O$ is a symmetry operation of the system, the response should be invariant,
$\tilde \sigma = \sigma$ and $\tilde \chi = \chi$, which leads to the symmetry
constraints
\begin{subequations}
\label{app_q}
\begin{gather}
\sigma_{\alpha\beta} = O_{\alpha m} O_{\beta n} \sigma_{mn}, \\
\chi_{\alpha\beta\gamma} = O_{\alpha m} O_{\beta n} O_{\gamma p} \chi_{mnp}.
\end{gather}
\end{subequations}
Plugging Eq.\,\eqref{app_o} into Eqs.\,\eqref{app_q} yields the
$\mathcal{C}_{3v}$-symmetry constraints on $\sigma_{\alpha\beta}$,
$\chi_{\alpha\beta\gamma}$, Eqs.\,\eqref{eq_4} of the main text.

\begin{table}
\caption{\label{tab1}The point group $\mathcal{C}_{3v}$.}
\begin{ruledtabular}
\begin{tabular}{l|llllll}
& $E$ & $C_3$ & $C_3^2$ & $\s_v'$ & $\s_v''$ & $\s_v'''$ \\ \hline
$E$ & $E$ & $C_3$ & $C_3^2$ & $\s_v'$ & $\s_v''$ & $\s_v'''$ \\
$C_3$ & $C_3$ & $C_3^2$ & $E$ & $\s_v''$ & $\s_v'''$ & $\s_v'$ \\
$C_3^2$ & $C_3^2$ & $E$ & $C_3$ & $\s_v'''$ & $\s_v'$ & $\s_v''$ \\
$\s_v'$ & $\s_v'$ & $\s_v'''$ & $\s_v''$ & $E$ & $C_3^2$ & $C_3$ \\
$\s_v''$ & $\s_v''$ & $\s_v'$ & $\s_v'''$ & $C_3$ & $E$ & $C_3^2$ \\
$\s_v'''$ & $\s_v'''$ & $\s_v''$ & $\s_v'$ & $C_3^2$ & $C_3$ & $E$
\end{tabular}
\end{ruledtabular}
\end{table}

\subsection*{Failure of power-series expansion}

\begin{figure}
\includegraphics[width=\columnwidth]{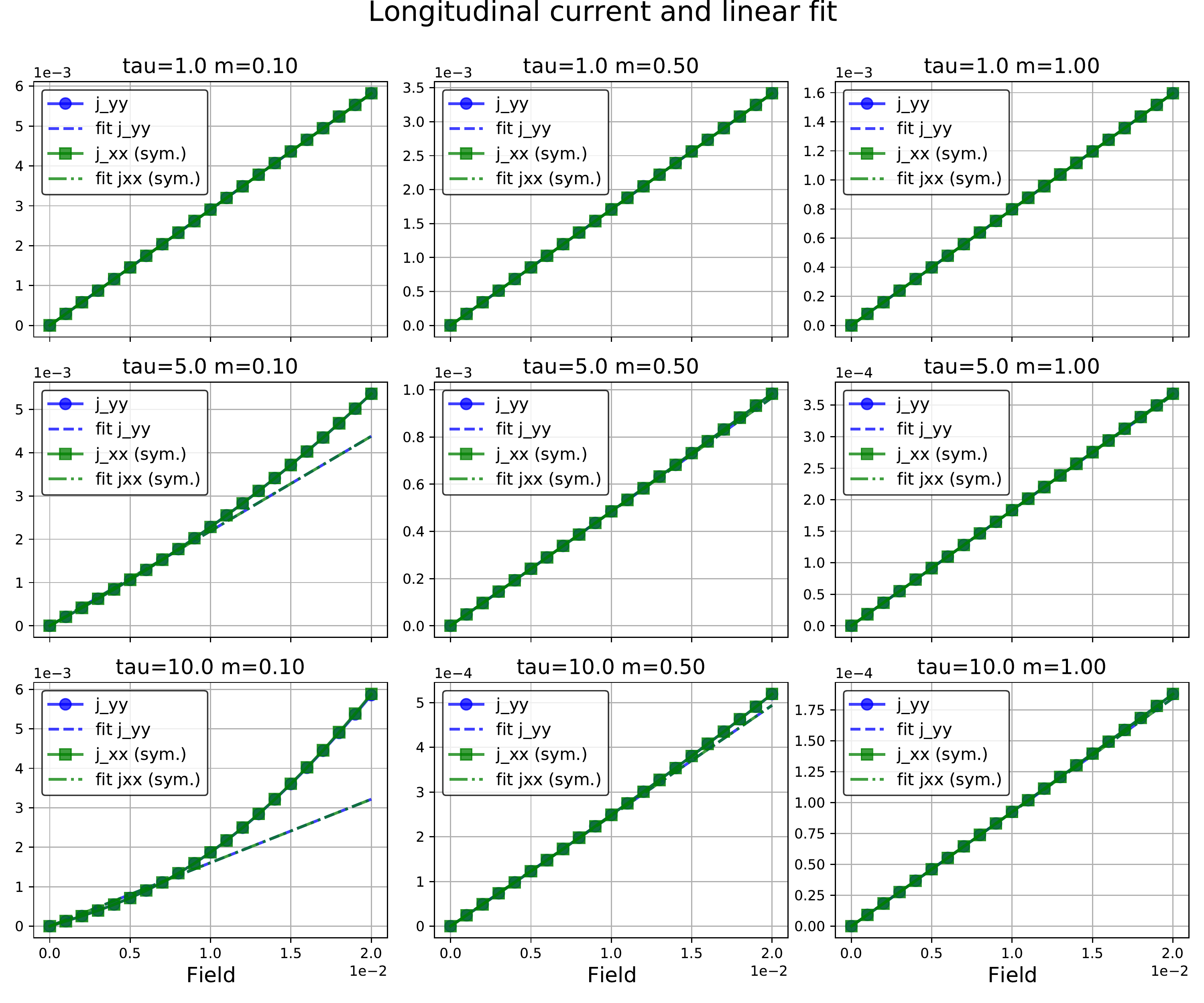}
\caption{\label{app_5}Steady-state longitudinal current $j_y(\bm E \pa \hat y)$
(blue circles) and symmetrized longitudinal current $j_x(\bm E \pa \pm \hat x)$
[see Eqs.\,\eqref{eq_6a}] (green squares) with linear fits (dashed and
dash-dotted) versus field amplitude for varying $\tau$ and $m$. Current
in units $et_h/(\hbar a)$, field in units $t_h/(ea)$.}
\end{figure}

\begin{figure}
\includegraphics[width=\columnwidth]{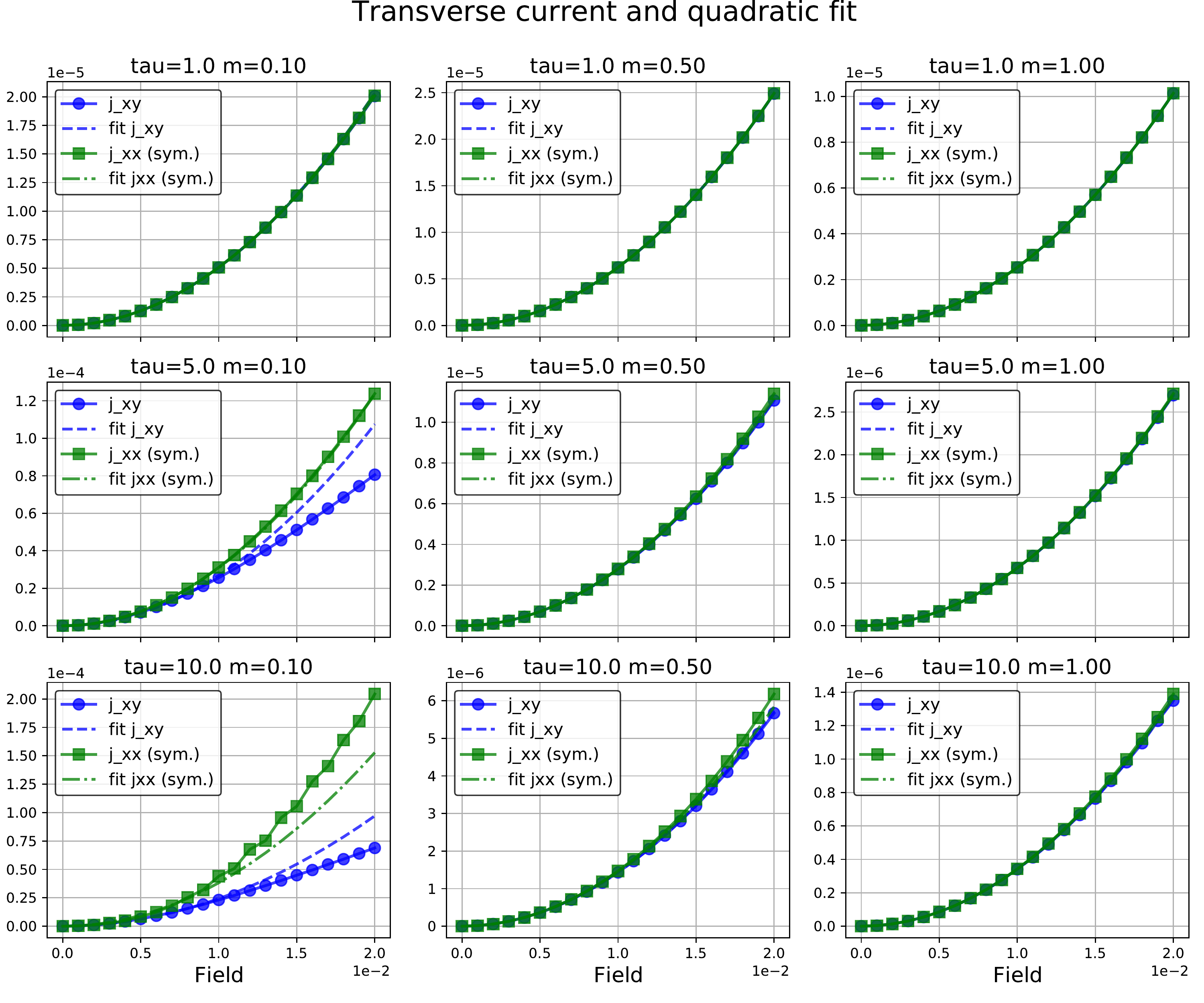}
\caption{\label{app_6}Steady-state transverse current $j_x(\bm E \pa \hat y)$
(blue circles) and symmetrized longitudinal current $j_x(\bm E \pa \pm \hat x)$
[see Eqs.\,\eqref{eq_6b}] (green squares) with quadratic fits (dashed and
dash-dotted) versus field amplitude for varying $\tau$ and $m$. Current
in units $et_h/(\hbar a)$, field in units $t_h/(ea)$.}
\end{figure}

In Fig.\,\ref{app_5} we plot the steady-state currents $j_y(\bm E \pa \hat y)$
and symmetrized $j_x(\bm E \pa \pm \hat x)$ cf.\ Eq.\,\eqref{eq_6a} together
with linear fits, similar to Fig.\,\ref{fig_2}(c). Similarly, in
Fig.\,\ref{app_6} we plot the steady-state currents $j_x(\bm E \pa \hat y)$ and
symmetrized $j_x(\bm E \pa \pm \hat x)$ cf.\ Eq.\,\eqref{eq_6b} together with
quadratic fits, similar to Fig.\,\ref{fig_2}(d). Fixing a field-amplitude
range, the power-series expansion Eq.\,\eqref{eq_j} and the consequent
Eq.\,\eqref{eq_2} fail for small $m$ and large $\tau$. It is however possible,
for any choice of finite $m$ and $\tau$, to take small enough $E$.



\end{document}